\documentclass[useAMS,usenatbib]{mn2e}

\usepackage{graphicx}
\usepackage{mathrsfs}
\usepackage{natbib}
\usepackage{amssymb}
\usepackage{overpic}
\usepackage{xcolor}
\bibpunct[,]{(}{)}{;}{a}{}{,}

\def\totd{{\mathrm{d}}}
\def\sun{{\odot}}

\title[Interplay of disk wind and dynamical ejecta]{The interplay of disk 
wind and dynamical ejecta in the aftermath of neutron star - black hole mergers}
\author[Fern\'andez, Quataert, Schwab, Kasen, \& Rosswog]{Rodrigo Fern\'andez$^{1,2}$, 
        Eliot Quataert$^2$, Josiah Schwab$^{1,2}$, 
        Daniel Kasen$^{1,3}$,
        \newauthor Stephan Rosswog$^4$\\
\\
$^1$ Department of Physics, University of California, Berkeley, CA 94720, USA.\\
$^2$ Department of Astronomy \& Theoretical Astrophysics Center, University of California, Berkeley, CA 94720, USA.\\
$^3$ Nuclear Science Division, Lawrence Berkeley National Laboratory, Berkeley, CA 94720, USA.\\
$^4$ The Oskar Klein Centre, Department of Astronomy, AlbaNova, Stockholm University, SE-106 91 Stockholm, Sweden.
}

\begin{document}

\date{Submitted to MNRAS}
\pagerange{\pageref{firstpage}--\pageref{lastpage}} 
\pubyear{2013}
\maketitle
\label{firstpage}

\begin{abstract}
We explore the evolution of the different ejecta components
generated during the merger of a neutron star (NS) and a black hole (BH). Our focus is
the interplay between material ejected dynamically during the merger, and the wind launched on 
a viscous timescale by the remnant accretion disk. These components are expected 
to contribute to an electromagnetic transient and to produce $r$-process elements,
each with a different signature when considered separately.
Here we introduce a two-step approach to investigate their combined evolution,
using two- and three-dimensional hydrodynamic simulations. Starting from
the output of a merger simulation, we identify each component in the initial condition
based on its phase space
distribution, and evolve the accretion disk in axisymmetry. The
wind blown from this disk is injected into a three-dimensional computational
domain where the dynamical ejecta is evolved.
We find that the wind can suppress fallback accretion on timescales
longer than $\sim 100$~ms. Due to self-similar viscous evolution, the 
disk accretion at late times nevertheless 
approaches a power-law time dependence $\propto t^{-2.2}$. This can power
some late-time GRB engine activity, although the available energy is significantly
less than in traditional fallback models.
Inclusion of radioactive heating due to the $r$-process does not significantly affect
the fallback accretion rate or the disk wind. We do not find any significant modification
to the wind properties at large radius due to interaction with the dynamical ejecta.
This is a consequence of the different expansion velocities of the two components.
\end{abstract}


\maketitle

\section{Introduction}

Double neutron star (NS) or NS - black hole (BH) mergers are among the
main candidates for direct detection of gravitational waves by 
ground based interferometers such as Advanced LIGO, Virgo, and 
KAGRA (e.g., \citealt{Abadie+10}). The ejecta from these mergers
is also a prime candidate for the generation of $r$-process elements 
(e.g., \citealt{Lattimer&Schramm74,Freiburghaus+99,goriely2013}).
Radioactive decay of these elements is expected to power an 
electromagnetic counterpart that can aid in the localization of 
the gravitational wave source \citep{Li&Paczynski98,Metzger+10b}. 
Finally, these mergers are the leading candidate progenitor for short-duration 
gamma-ray bursts (SGRBs; see, e.g., \citealt{berger2014} for a recent review).

The ejection of material by tidal forces during the merger depends
on a number of factors, including the mass ratio of the binary, the
spin of each component, and the properties of the nuclear equation of state 
(e.g., \citealt{bauswein2013,kyutoku2013}). In BH-NS mergers, the amount of ejecta also 
depends crucially on BH spin (e.g., \citealt{Foucart12}).
The gravitationally-bound part of this \emph{dynamical ejecta} leads to 
fallback accretion onto the BH. Because fallback extends
over timescales much longer than the viscous time of the disk,
it has been proposed as a source of extended prompt emission and/or 
X-ray flares in the afterglows of some short gamma-ray 
bursts \citep{Faber+06b,Gehrels+06,rosswog07}. 
However, the gravitational binding energy of material accreting on
timescales longer than $\sim 1$~s is comparable to or less than
the energy injected by radioactive heating during $r$-process nucleosynthesis.
Thus fallback accretion can potentially be suppressed when this heating is taken into account \citep{Metzger+10a}.

Furthermore, the dynamical ejecta does not exist in isolation. The accretion
disk formed during the merger is an additional source of material in two
ways. First, on $\sim 100$~ms timescales, material can be unbound by neutrino 
energy deposition in a broad polar outflow (the \emph{neutrino-driven
wind}, \citealt{McLaughlin&Surman05,surman2006,Surman+08,Dessart+09,Wanajo&Janka12,
MF14, perego2014}). Second, over longer, $\sim 1$~s timescales,
energy deposition by angular momentum transport and nuclear recombination
together with decreased neutrino cooling lead to substantial mass
ejection in a quasi-spherical outflow (a \emph{freezout wind}, 
\citealt{Metzger+09a,Lee+09,FM13,MF14,just2014,FKMQ14}). The amount of material
ejected by these two disk channels can be comparable to that in the 
dynamical ejecta, although its composition is expected to be less
neutron rich, with observable consequences for the ensuing kilonova transient 
(\citealt{Kasen+13,Barnes&Kasen13,tanaka2013,tanaka2014}, see also \citealt*{kasen2014}).

While the dynamical ejecta is generally launched earlier and
is faster than disk outflows, the interaction between
these two components can be non-trivial. First, disk winds can
modify fallback accretion relative to what is expected from purely
ballistic trajectories.
Conversely, part or all of the disk wind can mix with slower-moving 
components of the dynamical ejecta, potentially leading to different
nucleosynthetic and electromagnetic signatures than
predicted from the wind alone.

In this paper we examine the interplay between these components
by means of two- and three-dimensional hydrodynamic simulations.
Given that evolving the complete system over the timescales
of interest with all the relevant physics is computationally expensive,
we develop a two-stage modeling approach that takes advantage of 
the spatial and temporal decoupling of key processes.
Starting from the output of a BH+NS merger simulation, we evolve the 
remnant accretion disk
in axisymmetry (2D), with and without the dynamical ejecta, and 
including neutrino and viscous processes. 
This leads to the production of an axisymmetric disk wind. This wind
is then sampled at a radius that approximately separates the 
accretion disk from the non-axisymmetric dynamical ejecta.
The sampled wind is then injected into a larger, three-dimensional (3D) 
computational domain with an inner boundary that coincides with
the wind sampling radius, and in which the dynamical ejecta is evolved.
Our goal is to identify the
key processes that govern the interaction between these two
ejecta components using approximate modeling of the physics. 
Future studies will refine this analysis.

The paper is organized as follows. Section~\ref{s:methods} describes
the numerical approach employed, including the separation of the
system into different ejecta components. Section~\ref{s:results}
presents our results, including the properties of the disk wind in 2D, 
the properties of fallback accretion without wind,
and the interplay between these two components. Our results are summarized in
Section~\ref{s:summary}, where broader observational implications 
are also discussed.

\section{Computational Methodology}
\label{s:methods}

It is infeasible to evolve the combined
accretion disk plus dynamical ejecta system with all
the relevant physics in 3D for the timescales of interest ($\sim 10$~s).
In particular, explicit viscous angular momentum transport
requires resolving the corresponding diffusion time scale
on the smallest grid cells in the simulation. 

Fortunately, the (non-axisymmetric) dynamical ejecta is mostly
spatially decoupled from the (largely axisymmetric) 
accretion disk once the initial transient phase ends, a few dynamical times after the
merger. Angular momentum transport and neutrino processes
operate on timescales slower than the expansion time outside
radii $\sim 10^8$~cm (\S\ref{s:wind_injection}), so they
can be neglected in these regions to first approximation.

We thus adopt a two-stage approach to model the interaction
of wind and dynamical ejecta. First, we evolve the system
in axisymmetry including viscous and neutrino source terms, and measure
the properties of the
resulting disk wind at a radius that separates the 
two components. We then evolve
the dynamical ejecta in 3D, using a computational domain that 
has its inner radial boundary at the radius where the 
wind properties were determined. From this inner boundary, the wind is 
injected into the domain. The only source terms employed
in this second step are gravity and radioactive heating, leading to
significant savings in computational time.

In what follows, we describe the numerical method used
to evolve the hydrodynamic equations (\S\ref{s:hydro_method}),
the initial conditions, including how we identify
and separate distinct components of the merger remnant (\S\ref{s:initial_condition}),
the sampling and injection of the wind in 3D models (\S\ref{s:wind_injection}),
and the list of models simulated (\S\ref{s:models_evolved}).

\subsection{Time-dependent hydrodynamics}
\label{s:hydro_method}

The equations of hydrodynamics are solved numerically
with the dimensionally-split PPM solver in {\small \textsc{FLASH3.2}} 
\citep{fryxell00,dubey2009}. The public version
of the code has been modified to include a non-uniformly
spaced grid \citep{F12}, and physical source terms 
needed to model the evolution of merger remnant
accretion disks \citep{FM12,FM13,MF14,FKMQ14}. The equation of
state is that of \citet{timmes2000}, with abundances
of neutrons, protons, and alpha particles in nuclear
statistical equilibrium. The nuclear
binding energy contribution from alpha particles is included.

\begin{figure*}
\includegraphics*[width=\columnwidth]{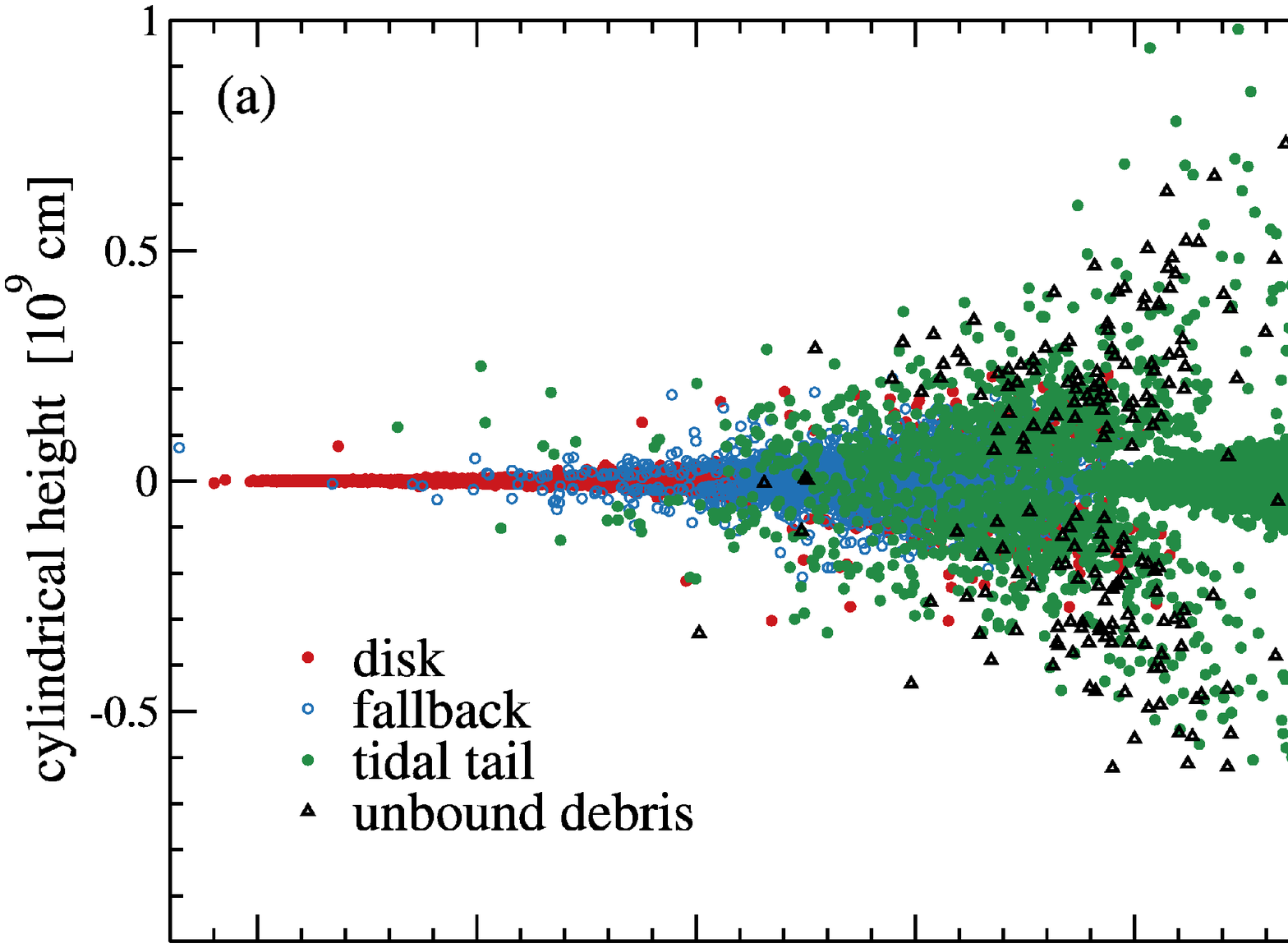}
\includegraphics*[width=\columnwidth]{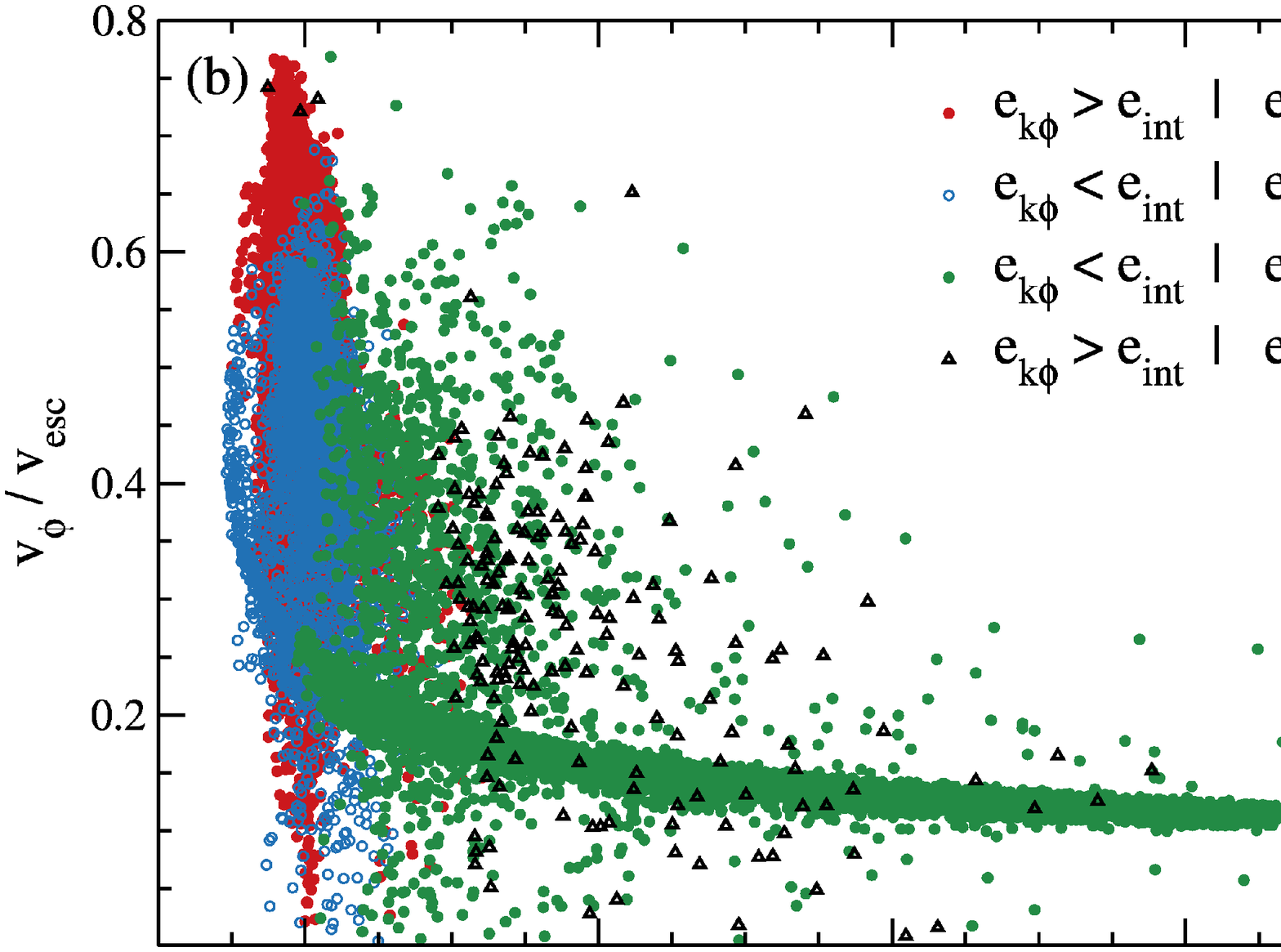}
\includegraphics*[width=\columnwidth]{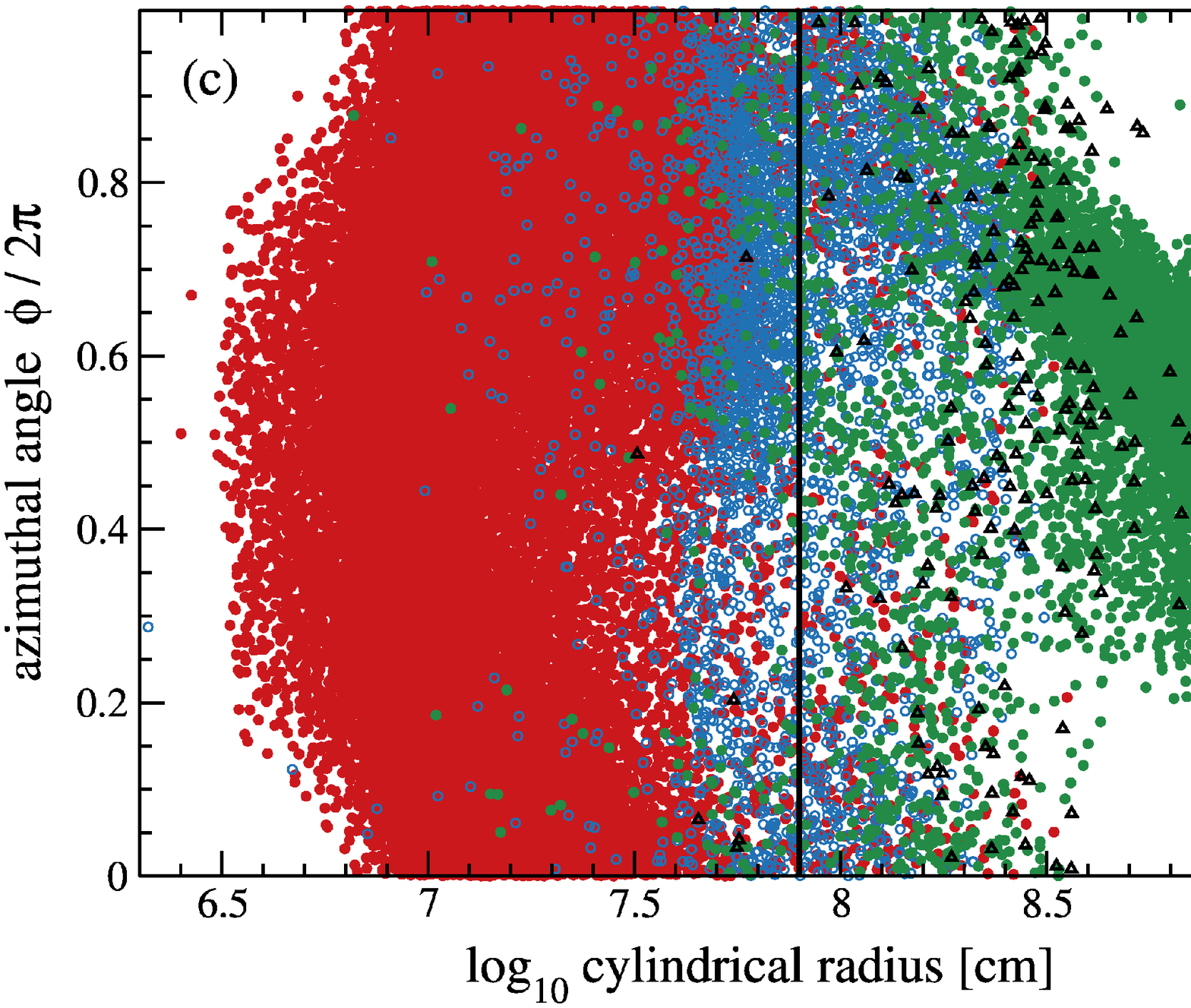}
\includegraphics*[width=\columnwidth]{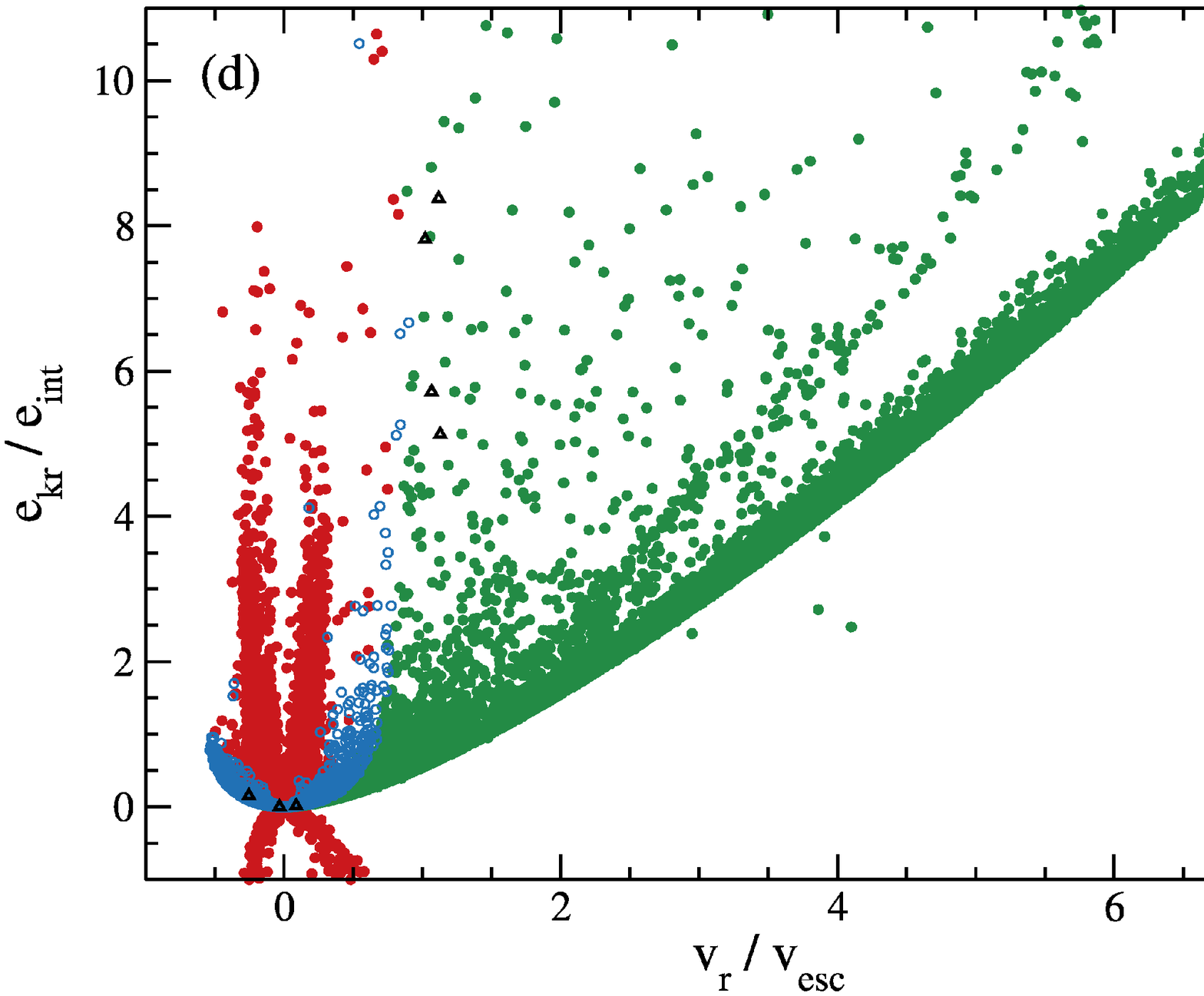}
\caption{Phase space distribution of particles that comprise
the initial state of the system. The disk (red solid circles), fallback (blue open circles),
tidal tail (green filled circles), and unbound debris (black triangles) components are separated
according to the sign of their total specific energy $e_{\rm tot}$, and
the ratio of the kinetic energy in the $\phi$ direction
$e_{\rm k\phi}$ to the internal energy $e_{\rm int}$, as indicated
in panel (b). The vertical line in panel (c) indicates the
radius $r_{\rm cut}$ at which the disk wind properties are measured
from 2D simulations. This radius marks the inner radial boundary
of 3D models.}
\label{f:particle_mapping}
\end{figure*}

The two-dimensional (2D) version of the code solves
the equations of mass-, poloidal momentum-, angular momentum-, 
energy-, and lepton number conservation in spherical polar
coordinates $(r,\theta)$. Source terms include
angular momentum transport by an anomalous shear stress using the kinematic
viscosity of \citet{shakura1973}, and neutrino source
terms via a leakage scheme that only includes
charge-current weak interactions \citep{MF14}.

The three-dimensional (3D) implementation solves
the equations of hydrodynamics in spherical polar
coordinates $(r,\theta,\phi)$ with gravity and
radioactive heating as the only source terms. 
The split PPM version of {\small {\textsc FLASH3.2}}
requires minor modifications to be extended to 3D spherical 
coordinates. A description of these modifications and
tests of the implementation will be presented elsewhere 
(Fern\'andez 2015, in preparation).
We employ the analytic parameterization of $r$-process radioactive heating
from \citet{Korobkin+12}. This source term is included in order to
qualitatively assess the effect of radioactive heating 
on the dynamics; details such as its sensitivity
to the electron fraction and spatial location are ignored.
This source term is applied in cells that have temperature $T < 5\times 10^9$~K. 

In all models, we approximate the gravitational potential of a
spinning BH via the pseudo-Newtonian potential of \citet{artemova1996}.
The main advantage of this potential is that it reproduces the location of 
the innermost stable circular orbit (ISCO) of the Kerr metric, 
and leads to steady-state, thin, 
sub-Eddington accretion disk solutions to within $\sim 10-20\%$ of the exact 
relativistic value as computed by \citet{artemova1996}. The implied spacetime is
spherically symmetric, however, so while convenient for computational purposes,
it is not a very accurate approximation for all space if the spin parameter is high.
Nevertheless, since most of the material that resides at radii close to the BH --
where GR effects are the strongest --
lies mostly on the midplane, we consider
this choice of potential as an acceptable approximation for the inner disk dynamics. At the
radii where the phenomena we are interested in occur ($r \gtrsim 10^8$~cm), 
GR effects are very weak (few percent or less).

The computational domain is discretized 
logarithmically in radius, uniformly in
$\cos\theta$ along the polar direction, and 
uniformly in azimuth. The resolution is
$64$ cells per decade in radius, $56$ cells meridionally from
$\theta = 0$ to $\theta = \pi$, and $192$ cells for $\phi \in [0,2\pi]$.
At the equator, cells are such that 
$\Delta r / r \simeq \Delta \theta \simeq \Delta \phi \simeq 2^\circ$.
The $(r,\theta)$ resolution is the same for most 2D and 3D models,
the exception being a 2D model run at double resolution to check for
convergence.

In 2D models that include the inner disk evolution, the domain covers
the full range of polar angles, and extends from a radius halfway 
between the BH horizon and the innermost stable circular orbit (ISCO),
until a radius $1000$ times larger. The radial limits of 3D models
are the wind injection radius $r_{\rm cut}$ (\S\ref{s:initial_condition}) on 
the inside, and a radius $10^4$ times larger on the outside.

The boundary conditions are reflecting in $\theta$ and periodic
in $\phi$. In 2D models and 3D models with no wind injection, both radial 
boundaries allow mass to leave the domain (see, e.g., \citealt{FM13} for details).
When a wind is injected into a 3D model, the default boundary condition involves
solution of a Riemann problem. This procedure is discussed in \S\ref{s:wind_injection}.

\subsection{Initial Condition and Separation of Components}
\label{s:initial_condition}

The initial condition is obtained from the output of 
a Newtonian Smoothed Particle Hydrodynamic (SPH) simulation
of the merger of a $10M_\sun$ BH with a $1.4M_\sun$ NS 
reported in \citet{Rosswog+12}. This BH mass is close to (but somewhat
higher than) the peak of the inferred stellar mass BH distribution
in the galaxy (e.g. \citealt{ozel2010,farr2011}).
The simulations use the
\citet{shen1998} equation of state,
a multi-flavor, energy-integrated neutrino leakage scheme \citep{Rosswog&Liebendorfer03}, and
Newtonian gravity with an absorbing boundary condition 
at the Schwarzschild radius of the (non-spinning) BH.
The time of the snapshot corresponds to $139$~ms after the
merger.

The SPH data is mapped into the Eulerian grid
using the appropriate smoothing kernel to
reconstruct conserved quantities from the particle
distribution. In the case of 2D simulations, data is 
axisymmetrized by computing azimuthal 
averages of conserved quantities (e.g., the radial velocity
is given by $\langle v_r\rangle = \int (\rho v_r)\totd V / \int \rho \totd V$,
with $\totd V$ the volume of the cells included in the average). We use
the density, temperature, and electron fraction to reconstruct
the rest of the thermodynamic variables using our equation of state.
If the temperature is below $5\times 10^9$~K, we assume full recombination
into alpha particles (e.g. if $Y_e < 0.5$, the alpha mass fraction is set to 
$2Y_e$ and the proton fraction to zero).

\begin{table*}
\centering
\begin{minipage}{17.5cm}
\caption{Models evolved and summary of results. The first six models follow the evolution
of the system in 2D, including all components (C2d, C2d-res, C2d-src, C2d-h), material
interior to $800$~km (C2d-df) or just disk material (C2d-d). 
The second four models explore the properties of fallback accretion 
in 2D (F2d, F2d-h) and 3D (F3d, F3d-h) neglecting the disk wind. 
The final five models investigate the effect of
injecting the disk wind measured in 2D simulations into a domain containing the
dynamical ejecta in 2D (I2d) and 3D (I3d, I3d-nR, I3d-df, I3d-h). 
The first eight columns from the left show model name, dimensionality, position of the 
inner and outer radial boundaries, use of viscous and neutrino source terms, 
injection of disk wind from the inner radial boundary,
inclusion of radioactive heating, and type of inner radial 
boundary condition (out: outflow, rmn: riemann solver, inf: inflow), 
respectively\label{t:models}. The following two columns show
the total and unbound mass ejected in wind material 
(\emph{disk} and \emph{fallback} components) at $r=10^9$~cm within $10$s, respectively, while the
final two columns show the mass-flux-weighted, time-averaged
radial velocity and electron fraction of the unbound wind component at
$r=10^9$~cm, respectively. 
The black hole
has a mass $M_{\rm bh}=11.1M_\sun$ and spin parameter $a=0.8$ in all cases. See \S\ref{s:models_evolved}
for other parameters.}
\begin{tabular}{lccccccccccc}
\hline
{Model} & 
{Dim.} & 
{$r_{\rm min}$} &
{$r_{\rm max}$} & 
{Source} & 
{Wind} & 
{Rad.} & 
{Bnd.} &
{$M_{\rm w,t}$} & 
{$M_{\rm w,u}$} &
{$\bar{v}_{\rm w,u}$} & 
{$\bar{Y}_{\rm e, u}$} \\ 
{} & {} & \multicolumn{2}{c}{(km)} & {Terms} & {Inj.} & {Heat} & {Cnd.} & \multicolumn{2}{c}{($10^{-2} M_\sun$)} & {$(10^{-2}c)$} 
& {} \\
\hline
C2d         & 2D & 37  & 3.7E+4  & Y   & ...  & N   & out & 3.9 & 2.1 & 3.9 & 0.26 \\
C2d-res     &    &     &         &     & ...  &     &     & 3.6 & 1.7 & 3.8 & 0.26 \\
C2d-df      &    &     &         &     & ...  &     &     & 3.5 & 1.5 & 3.8 & 0.29 \\
C2d-d       &    &     &         &     & ...  &     &     & 3.3 & 1.6 & 4.2 & 0.29 \\
C2d-src$^a$ &    &     &         &     & ...  &     &     & 4.4 & 1.4 & 5.5 & 0.27 \\
C2d-h       &    &     &         &     & ...  & Y   &     & 4.1 & 2.4 & 4.3 & 0.27 \\
\noalign{\smallskip}
F2d    & 2D & 800 & 8E+6 & N   & N   & N   & out      & ... & ... & ... & ...  \\
F3d    & 3D &     &      &     &     &     &          & ... & ... & ... & ...  \\
F2d-h  & 2D &     &      &     &     & Y   &          & ... & ... & ... & ...  \\
F3d-h  & 3D &     &      &     &     &     &          & ... & ... & ... & ...  \\
\noalign{\smallskip}
I3d    & 3D & 800 & 8E+6 & N   & Y   & N   & rmn   & 1.8 & 1.4 & 5.4 & 0.28 \\
I3d-nR &    &     &      &     &     &     & inf   & 2.0 & 1.5 & 5.1 & 0.27 \\
I3d-df &    &     &      &     &     &     & rmn   & 1.0 & 0.8 & 5.7 & 0.28 \\
I3d-h  &    &     &      &     &     & Y   &       & 2.5 & 1.9 & 4.9 & 0.27 \\
I2d    & 2D &     &      &     &     &     &       & 1.5 & 1.2 & 7.0 & 0.28 \\
\hline
\noalign{\smallskip}
\noalign{$^a$ This model suppresses neutrino and viscous source terms outside $r=800$~km.}
\end{tabular}
\end{minipage}
\end{table*}

In order to separate the dynamical ejecta from the disk, 
we inspect the properties of the SPH particles in phase space
and identify distinctive features. Figures~\ref{f:particle_mapping}b and
\ref{f:particle_mapping}d show the kinematic distribution of particles relative to the 
gravitational and internal energies. Two distinct components are
evident, one that extends to large values of the radial velocity $v_r$ relative
to the escape speed $v_{\rm esc}$ (and hence moves ballistically), and another that clusters 
isotropically around zero radial velocity. The nature of these components 
becomes clear when they are sub-divided according to their degree 
of gravitational binding, quantified by
the sign of the total specific energy,
\begin{equation}
\label{eq:total_energy}
e_{\rm tot} = e_{\rm kr}+e_{\rm k\theta}+e_{\rm k\phi} + e_{\rm int}+e_{\rm grv},
\end{equation}
and by the degree of rotational support, quantified by the ratio of rotational-kinetic
to internal energies $e_{\rm k\phi}/e_{\rm int}$,
where $e_{{\rm k}i}$ is the kinetic energy along the $i$-th direction,
$e_{\rm int}$ is the internal energy, and $e_{\rm grv}$ is the gravitational energy,

We identify four components in the system:
\begin{enumerate}
\item \emph{Disk}: gravitationally bound ($e_{\rm tot}<0$) and rotationally 
	   supported ($e_{\rm k\phi}>e_{\rm int}$). This is the 
	   innermost component in radius and nearly axisymmetric,
	   as shown in Figures~\ref{f:particle_mapping}a and \ref{f:particle_mapping}c, 
	   containing most of the mass ($0.2M_\sun$).
	   \newline

\item \emph{Tidal tail}: gravitationally unbound and with $e_{\rm k\phi} < e_{\rm int}$.
		 The energy is dominated by the radial kinetic energy, as shown
		 in Figure~\ref{f:particle_mapping}, comprising a highly
		 localized structure in phase space, outermost in radius.
		 The mass in this component is $0.06M_\sun$.
		 \newline

\item \emph{Fallback}: gravitationally bound, and primarily gas pressure supported ($e_{\rm k\phi} < e_{\rm int}$).
	       With this definition, it includes the outermost part of the disk
	       and the innermost part of the tidal tail, as shown most explicitly
	       in Figure~\ref{f:particle_mapping}b. The mass in this component
	       is $0.02M_\sun$.
		\newline

\item \emph{Unbound debris}: the remaining material is gravitationally unbound with
		     significant rotation. It can be considered
		     part of the dynamical ejecta, but it does not comprise
		     as localized a structure as the tidal tail. The mass is $\sim 10^{-3}M_\sun$.
			\newline
\end{enumerate}
We follow the separate evolution of these components in our {\small\textsc FLASH} simulations using passive
scalars, constructed by taking the ratio of the partial mass density formed with each sub-group
of SPH particles to the total mass density. By definition, these scalars add up to unity.

Once mapped into {\small\textsc FLASH}, these components are not completely decoupled, 
however. Figures~\ref{f:particle_mapping}a 
and \ref{f:particle_mapping}c show that there is spatial overlap between the particles, 
particularly around $r\sim 10^8$~cm. This overlap leads to mixing when reconstructing the 
fluid distribution using the SPH kernel.
Approximately $0.01M_\sun$ of material initially tagged as \emph{tidal tail} is gravitationally 
bound, while a very small amount of \emph{disk} and \emph{fallback} material 
($\simeq 10^{-3}M_\sun$ in total) is unbound.

We choose a radius $r_{\rm cut} = 800\textrm{ km}$ to quantify the 
unbound properties of the disk wind in 2D axisymmetric simulations
of the inner disk. This position corresponds approximately
to the innermost edge of the tidal tail (Figure~\ref{f:particle_mapping}c), 
separating a nearly axisymmetric matter distribution on the inside from a highly
non-axisymmetric one on the outside. At the time of our initial conditions
($139$~ms after the merger), 
this surface encloses $99\%$ of the
\emph{disk} mass and $43\%$ of the \emph{fallback} mass.

The mass of the BH at the beginning of the simulation is
approximately $11.1M_\sun$, and it is kept fixed thereafter. The inner boundary for 2D 
models is set at $3.7\times 10^6$~cm, midway between the ISCO
and the event horizon of a BH with spin parameter $a = 0.8$.
A significantly smaller spin would be inconsistent with the
tidal disruption of the NS (e.g. \citealt{Foucart12}). Also,
the potential of \citet{artemova1996} asymptotes to Newtonian
for very high spins, so this choice is more consistent
with the physics used to generate the initial
condition (a BH modeled as a Newtonian point mass with an 
absorbing boundary condition at the Schwarzschild radius of
a non-spinning BH; \citealt{Rosswog+12}).

\subsection{Wind Injection in 3D Simulations}
\label{s:wind_injection}

In 2D simulations, we record the properties of the material
crossing at $r=r_{\rm cut}$, as a function of polar angle
and at regular time intervals. Subsequent injection of this data 
into 3D models is achieved by interpolating the recorded
variables in time for a given polar angle, copying the
resulting values for all azimuthal angles.

The inner radial ghost cells of 3D models are filled by solving 
an Riemann problem at the inner radial boundary, taking as
left state the quantities interpolated from the sampled
wind, and as right state the innermost active cell in radius.
This is done to account for the possibility that fallback
may carry a larger momentum flux than the wind along certain
directions. For operational simplicity, we employ the HLLC solver of \citet{toro1994},
which does not require any iterations. As a check, we run a model that simply copies the
interpolated values from the 2D simulations into the ghost cells of the 3D domain, 
without a Riemann solution.

In all cases, if the resulting radial velocity at the ghost
cells is negative, a standard outflow boundary condition
is adopted: the ghost cells are filled with data from the
innermost active cell.

\begin{figure*}
\includegraphics*[width=\textwidth]{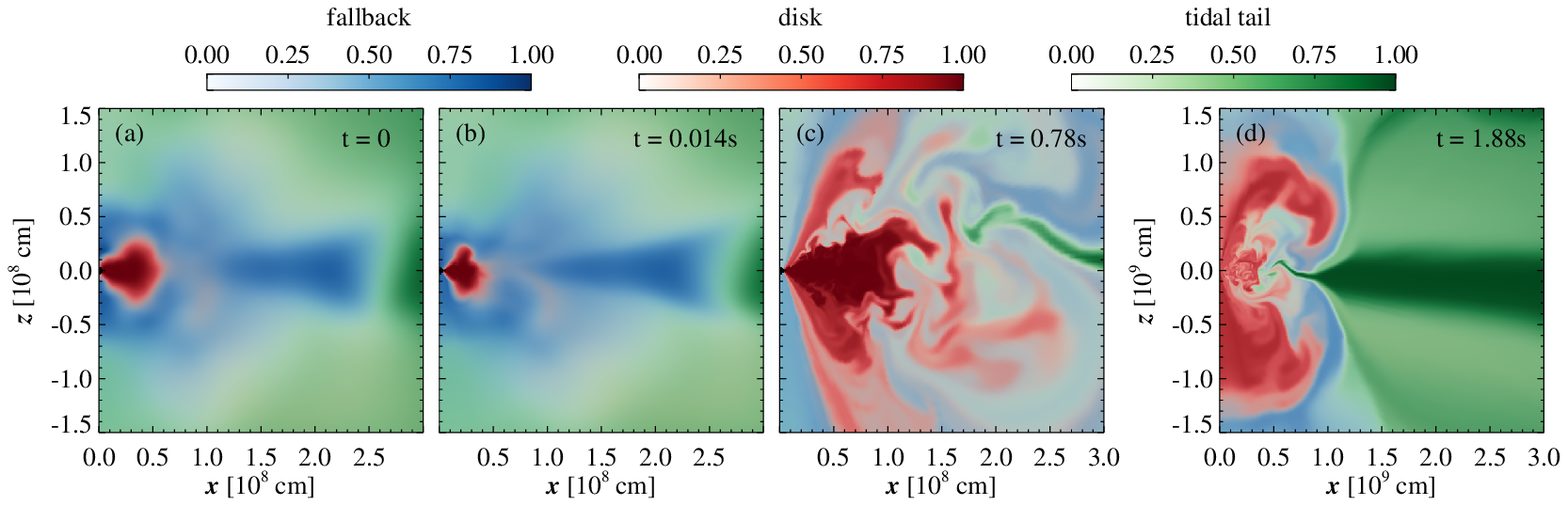}
\includegraphics*[width=\textwidth]{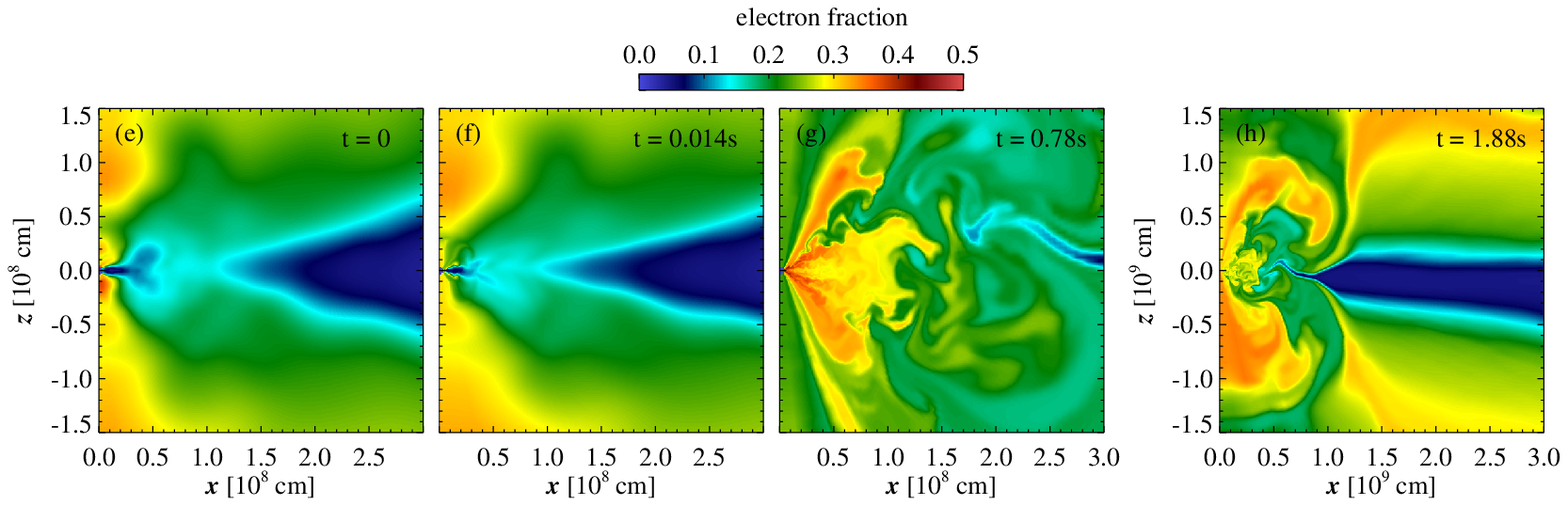}
\caption{Snapshots in the evolution of the high-resolution 2D model C2d-res,
which evolves the entire merger remnant (disk + dynamical ejecta) including the effects of
shear viscosity and neutrinos. 
The upper four panels (a-d) show the mass fractions of passive scalars that trace the
different components of the system as defined in \S\ref{s:initial_condition}:
disk (red), fallback (blue), and tidal tail (green). The lower four panels (e-h)
show electron fraction. Note the change in $x$- and $z$ scale in panels (d) and (h).}
\label{f:wind_2d_phases}
\end{figure*}

Outside $r=r_{\rm cut}$, the energy source terms that are important
for the evolution of the inner disk operate on timescales longer than 
the expansion time, and are therefore neglected in 3D simulations. To show this,
we explicitly evaluate these timescales. The
expansion time is
\begin{equation}
\label{eq:expansion_time}
t_{\rm exp} = \frac{r}{v_r} \simeq 0.1\, r_8\left(\frac{0.03c}{v_r} \right)~\textrm{ s},
\end{equation}
where $r_8 = r/10^8$~cm and $v_r\simeq 0.03c$ for the disk wind (e.g., \citealt{FKMQ14}). 
This is comparable to the free-fall time at this radius (for the dynamical ejecta, $v_r\gtrsim 0.1c$, which is 
essentially the Keplerian velocity for a $10M_\sun$ central mass). The
viscous time is
\begin{eqnarray}
t_{\rm visc} &\simeq & \frac{r^2}{\nu} = \frac{1}{\alpha}\frac{r}{v_{\rm K}} \left(\frac{H}{R} \right)^{-2}\nonumber\\
     &\simeq & 1\left(\frac{0.03}{\alpha} \right)\left(\frac{H}{R} \right)^{-2}M_{10}^{-1/2}r_8^{3/2}\textrm{ s},
\end{eqnarray}
where $\alpha$ is the viscosity parameter of \citet{shakura1973}, $(H/R)$ is the ratio of the disk scaleheight
to the local cylindrical radius, $v_{\rm K}$ is the Keplerian velocity, and $M_{10} = M_{\rm bh}/10M_\sun$.
The cooling time is
\begin{eqnarray}
t_{\rm cool} & =      & \frac{e_{\rm int}}{Q_\nu^{-}} \simeq \frac{(H/R)^2 v_{\rm K}^2}{Q_\nu^{-}}\nonumber\\
     & \simeq & 400 \left(\frac{H}{R} \right)^2 M_{10} r_8^{-1}\left(\frac{kT}{0.5\textrm{ MeV}}\right)^{-6}\textrm{ s},
\end{eqnarray}
where we have used an approximation to the charged-current weak interaction 
emissivity from \citet{janka2001},
$Q_\nu^{-} \simeq 145\left(T/2\textrm{ MeV}\right)^6$~MeV~s$^{-1}$ per baryon, where $T$ is the local gas temperature. 
Similarly, the neutrino heating time is
\begin{eqnarray}
t_{\rm heat} & \simeq & \frac{(H/R)^2 v_{\rm K}^2}{Q_\nu^{+}}\nonumber\\
     & \simeq & 1\left(\frac{H}{R} \right)^2 M_{10} r_8 T_{\nu,4}^{-2} L_{\nu,53}^{-1}\textrm{ s},
\end{eqnarray}
where the approximation for charged-current neutrino absorption by nucleons in \citet{janka2001} 
has been used, $Q_\nu^{+} \simeq 16 L_{\nu,53} T_{\nu,4} r_7^{-2}$~MeV~$s^{-1}$ per baryon.
Here $T_{\nu,4}$ is the neutrinosphere temperature in units of $4$MeV, and $L_{\nu,53}$ is the
neutrino luminosity in units of $10^{53}$~erg~s$^{-1}$. Finally, the radioactive heating
timescale is
\begin{eqnarray}
t_{\rm rad} \simeq \frac{e_{\rm int}}{\dot{\epsilon}_0} \simeq 7\left(\frac{H}{R} \right)^2 M_{10} r_8^{-1}\textrm{ s},
\end{eqnarray}
where $\dot{\epsilon}_0 = 2\times 10^{18}$~erg/(g s) is the amplitude of the radioactive
heating fit of \citet{Korobkin+12}. 

The hierarchy of timescales at $r \sim r_{\rm cut}$ is therefore
\begin{equation}
t_{\rm exp} \ll t_{\rm rad}\sim t_{\rm heat}\sim t_{\rm visc} \ll t_{\rm cool}.
\end{equation}
Both $t_{\rm heat}$ and $t_{\rm visc}$ increase as the material expands, whereas
$t_{\rm rad}$ decreases with increasing radius, at least during the first second
of evolution where the heating rate remains nearly constant. This motivates us
to include radioactive heating while neglectic neutrino and viscous source terms
in 3D simulations covering the region $r > r_{\rm cut}$. 
While this approximation is strictly valid only for 
times $t \lesssim t_{\rm visc}(r_{\rm cut})\sim 1$~s, we adopt it for all times
as a first approximation to explore the behavior of the system.

\subsection{Models evolved}
\label{s:models_evolved}

Table~\ref{t:models} summarizes our simulations. We run
three sets: one that evolves the complete system in 2D with all the physics
(C-series, for ``complete"), one that investigates 
fallback accretion without wind (F-series, for ``fallback"), 
and one that evolves the dynamical ejecta with wind
injection from the inner radial boundary (I-series, for ``injection").

The fiducial 2D model (C2d) includes all the axisymmetrized components
of the system. This model is repeated at twice the resolution in radius
and angle to test for convergence (C2d-res).
The next two models test the properties of the disk and wind
when excluding all material outside $r=r_{\rm cut}$ in the 
initial condition $r_{\rm cut}$ (C2d-df), or including only material
tagged initially as \emph{disk} (C2d-d; \S\ref{s:initial_condition}).
The final two models explore how the system changes when
including radioactive heating (C2d-h) and suppressing all source
terms outside of $r=r_{\rm cut}$ (C2d-src).
 
Four additional models explore the properties of fallback accretion
without wind injection. Two models follow the evolution
of all material outside $r=r_{\rm cut}$ without source
terms other than gravity, one in 2D and one in 3D 
(F2d and F3d, respectively). Two additional models
repeat this calculation, adding radioactive heating
(F2d-h and F3d-h).

The last five models explore the interaction of the
tidal tail with wind material injected at $r=r_{\rm cut}$. 
The fiducial 3D model (I3d) uses the wind from model
C2d and employs a Riemann boundary condition for injection.
A second model (I3d-nR) employs a simpler boundary condition, in which
the interpolated wind variables are copied to the
ghost cells.
A third model (I3d-df) uses the wind sampled from model C2d-df (no tidal
tail) and solves a Riemann problem at the boundary.
The fourth model (I3d-h) repeats the fiducial model but now
adding radioactive heating. Finally, model
I2d is the same as I3d but in 2D, for comparison.

\section{Results}
\label{s:results}

\subsection{Disk evolution in axisymmetry}
\label{s:2d_evolution}

Disks formed in mergers involving neutron stars
undergo characteristic evolutionary phases determined by
the degree of neutrino cooling
(e.g. \citealt{popham1999,Narayan+01,Chen&Beloborodov07,FM13}).
Neutrino processes are initially important given the
high density and temperature of the torus ($\sim 10^{11}$g~cm$^{-3}$ and
$\sim 5$MeV, respectively). 
Despite the high torus mass ($0.2M_\sun$), the disk is not too optically thick
initially because this mass is spread over a relatively
large radial extent (the density maximum is located at $\sim 80$~km).
Given that the initial condition is not in equilibrium
and that the microphysics is not the same as that used in
the original merger simulation, the system displays a transient 
phase during the initial $\sim 0.02$s (several orbits
at the density maximum), adjusting to a new equilibrium state thereafter. 
The duration of this transient is much shorter than the timescale over which the phenomena 
we are interested in occur.

\begin{figure}
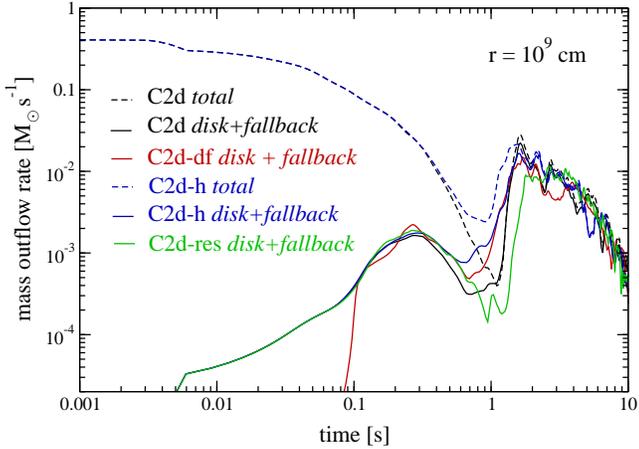

\begin{overpic}[width=\columnwidth,clip=true]{f3.eps}
\put(3.5,51.5){\tiny $\odot$}
\end{overpic}
\caption{Mass outflow rate at $r=10^9$~cm as a function of time for
different 2D models. The total outflow rate, including the tidal
tail, is shown by dashed lines. The `wind' (solid lines) includes
material that is tagged as \emph{disk} and \emph{fallback} in the initial condition, as
defined in \S\ref{s:initial_condition}. The qualitative evolution
relative to the baseline model (black) is independent of whether radioactive 
heating is included (blue), whether the tidal tail is excluded (red),
or if the resolution is doubled in radius and polar angle (green).}
\label{f:mdot_largeradius}
\end{figure}

Throughout the disk evolution, the contribution of neutrino absorption to
the overall heating rate at radii $r > 100$~km
is at most a few per cent of the viscous energy deposition
minus neutrino cooling (see also \citealt{FKMQ14}). We do not see a neutrino-driven
wind in our 2D models. Once the disk has spread sufficiently for 
its temperature and density to drop below values where neutrino cooling 
becomes inefficient (e.g., \citealt{Metzger+09a}), a viscously-driven 
outflow is launched. This occurs around $t \sim 1$~s.

Figure~\ref{f:wind_2d_phases} shows how the
different components of the system, as traced by passive scalars (\S\ref{s:initial_condition}),
interact during these evolutionary phases. After a few orbits at the
density maximum ($t = 0.014$s), during the
initial transient phase, the disk reaches a minimum size, presumably
due to accretion and compression by fallback material.
Once the disk expands due to angular momentum transport and becomes convective, 
a wind is launched, mixing the original \emph{disk} material with 
\emph{fallback} and \emph{tidal tail} matter ($t=0.78$~s). At late times, 
the wind expands primarily towards mid-latitude directions, away
from the midplane occupied by the tidal tail ($t= 1.88$~s).
Figure~\ref{f:wind_2d_phases} also shows the electron fraction
of the different components, contrasting the neutron-rich tidal
tail with the higher-$Y_e$ disk wind.

The mass outflow rate as a function of time at a radius of $10^9$~cm is 
shown in Figure~\ref{f:mdot_largeradius} for models C2d (all components), 
C2d-df (no tidal tail), 
C2d-h (radioactive heating), and C2d-res (all components at 
double resolution in radius and polar angle).
The contribution from the \emph{disk} and \emph{fallback} scalars is
shown as well as the total mass outflow including tidal tail material.
At early times, mass ejection is dominated by the tidal
tail, transitioning around $t\sim 1$~s to dominance
by the disk wind. At late times, the instantaneous outflow rate at large radii
is not very sensitive to whether the tidal
tail is present or not, and whether radioactive heating is
included. 

Table~\ref{t:models} shows integrated properties of the 2D models
over $10$~s of evolution. The total mass ejected is of the order of $20$ percent
of the initial disk mass ($\sim 0.04M_\odot$), in agreement with 
previous results for a BH with spin $a = 0.8$ \citep{just2014,FKMQ14}.
The fraction of this outflow that has positive specific energy (with
the internal energy normalized so that it vanishes at $T=0$) lies
between $50$ and $60$ percent. The mean velocity of this unbound
wind is $\sim 0.04c$. Radioactive heating leads to a $\sim 15$ percent
enhancement in the unbound mass ejection, whereas excluding
the tidal tail leads to $\sim 10$ percent less mass ejected due to the absence of
leftover material swept up by the wind. The overall uncertainty
due to resolution is $\sim 10$ percent. 

Quantities relevant for heavy-element nucleosynthesis are shown
in Table~\ref{t:wind_2d}. The mass-flux-weighted averages are
calculated according to \citet{FM13}:
\begin{equation}
\label{eq:average_definition}
\bar{A} = \frac{\int \totd t\,\totd \Omega\, F_{\rm M}(r_{\rm out},\hat{\Omega}) A(r_{\rm out},\hat\Omega)}
       {\int \totd t\,\totd \Omega\, F_{\rm M}(r_{\rm out},\hat{\Omega})},
\end{equation}
where $A$ is a generic quantity, $F_{\rm M} = \rho v_r$ is the mass flux, 
$\hat\Omega$ is the angular direction, and 
$r_{\rm out}\simeq 10^8$~cm is a radius chosen so that $\bar{T} \simeq 5\times 10^9$~K 
when computing the mean $Y_e$, entropy, and expansion time. In this case the angular range of the
integral is restricted to within $60^\circ$ of the midplane, because little material populates
the polar regions at these radii. In order to compare with 3D models, 
the average wind velocity and electron fraction shown in Table~\ref{t:models} are computed 
at $r = 10^9$~cm, including only material with positive specific energy, and including all angular directions.

\begin{table}
\centering
\begin{minipage}{8cm}
\caption{Mean properties of the disk wind in 2D models. Columns are model name,
electron fraction, entropy, and expansion time. The mass-flux-weighted,
time-averaged quantities are computed using equation~(\ref{eq:average_definition}),
at a radius where $\bar{T}\simeq 5\times 10^9$~K ($\sim 1000-1500$~km).
Only material tagged as \emph{disk} and \emph{fallback} is included.\label{t:wind_2d}}
\begin{tabular}{lccc}
\hline
{Model} & 
{$\bar{Y}_{\rm e}$} & 
{$\bar{s}$} &
{$\bar{t}_{\rm exp}$} \\
{} & {} & {(k$_{\rm B}/$b)} & {(ms)} \\
\hline
C2d      & 0.29 & 32 &  66 \\ 
C2d-df   & 0.27 & 29 & 111 \\ 
C2d-d    & 0.28 & 29 & 96  \\ 
C2d-h    & 0.28 & 32 &  83 \\ 
C2d-res  & 0.28 & 33 &  48 \\ 
C2d-src  & 0.27 & 29 &  28 \\ 
\hline
\end{tabular}
\end{minipage}
\end{table}

The mean electron fraction of the wind is $\sim 0.26-0.29$, slightly higher than that obtained
with a smaller BH mass and more compact disks, starting from an equilibrium initial condition,
and using the same neutrino implementation (e.g. \citealt{FKMQ14}). 
The mean entropies are $\sim 30k_{\rm B}$ per baryon and the
expansion time lies in the range $50-100$~ms. Given these parameters, the critical
electron fraction above which no lanthanides are produced is $\sim 0.25$ \citep{kasen2014}. 
The disk wind will therefore lead primarily to lanthanide-free material and a kilonova
component peaking in the optical band.

\begin{figure}
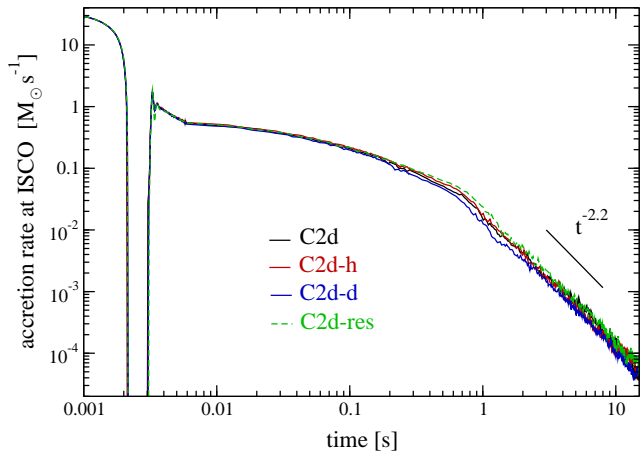

\begin{overpic}[width=\columnwidth,clip=true]{f4.eps}
\put(3.5,56){\tiny $\odot$}
\end{overpic}
\caption{Net mass accretion rate at the ISCO for 2D models (see Table~\ref{t:models}
for parameters). The late-time accretion rate is somewhat steeper than that due to 
ballistic fallback, and is set by the viscously spreading disk. The temporal
slope does not significantly depend on whether the dynamical ejecta 
and fallback are excluded, whether
radioactive heating is added, or on the resolution of the simulations.
At early times ($t\lesssim 0.01$~s) the disk is undergoing transient
readjustment.}
\label{f:mdot_isco}
\end{figure}

The accretion history at the ISCO is shown in 
Figure~\ref{f:mdot_isco}. The initial transient phase is evident,
with even a short period during which no accretion takes place.
From the end of this transient phase at $t\simeq 0.02$s until
approximately $t=1$s, the accretion rate evolves smoothly.
After the wind is launched, however, the time-dependence
of the accretion rate steepens, with the asymptotic power-law
in the range $t^{-2.15} - t^{-2.3}$. Such a drop in the accretion rate
at late times was first seen by \citet{Lee+09}. 
In the context of fallback accretion, we
note that this time-dependence is insensitive to whether material labeled as
\emph{fallback} and \emph{tidal tail} is included, as shown in Figure~\ref{f:mdot_isco}
(model C2d-d includes only material labeled as \emph{disk}).

Indeed, one can show using simple scaling arguments that during the late
radiatively-inefficient phase of evolution, the mass accretion rate 
scales like $t^{-4/3}$ \citep*{metzger2008steady}.
This scaling becomes even steeper ($t^{-8/3}$) when outflows are included in the solution.
Our models have an intermediate behavior between these two limits.

\begin{figure}
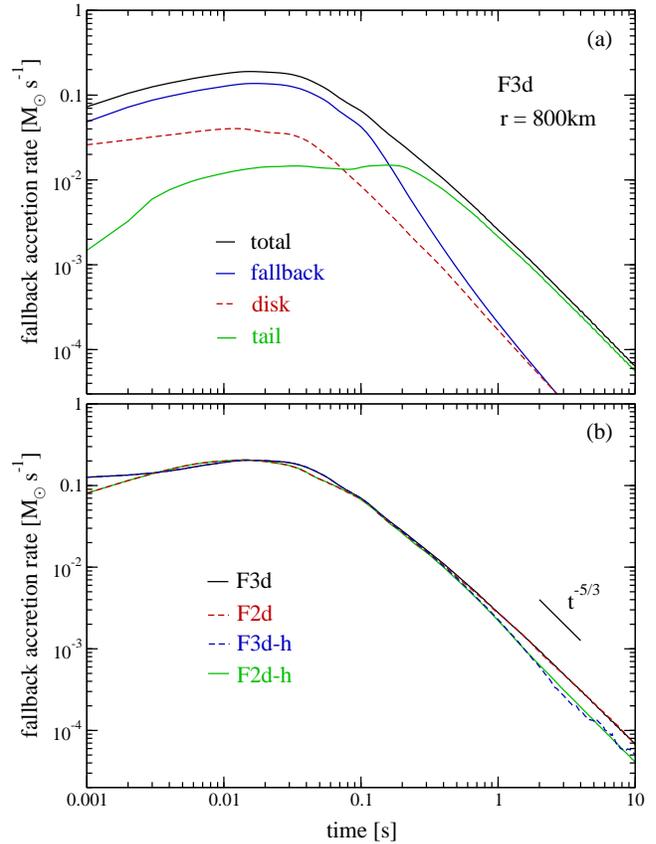

\begin{overpic}[width=\columnwidth,clip=true]{f5a.eps}
\put(3.5,46){\tiny $\odot$}
\end{overpic}
\begin{overpic}[width=\columnwidth,clip=true]{f5b.eps}
\put(3.5,54.5){\tiny $\odot$}
\end{overpic}
\caption{Mass accretion (fallback) rate as a function of time at $r = r_{\rm cut}=800$~km
for models without wind injection. Panel (a) shows the accretion rate for
model F3d, which ignores radioactive heating. Also shown are the contributions
from the different components of the system (\S\ref{s:initial_condition}) as traced by passive scalars.
Panel (b) compares 2D and 3D models with and without radioactive heating.
While radioactive heating causes a temporary
steepening of the accretion rate with time, its overall effect is small.
Models in 2D and 3D are very close to each other at late times.}
\label{f:fallback_dim}
\end{figure}

\subsection{Fallback accretion without disk wind: effect of radioactive heating}

If the gravitationally bound part of the dynamical
ejecta moves in Keplerian orbits, the expected
fallback accretion scales with time like (e.g., \citealt{rees1988})
\begin{equation}
\label{eq:mdot_fallback_definition}
\dot{M}_{\rm f} \propto \frac{\totd M}{\totd E_{\rm orb}} t^{-5/3},
\end{equation}
where $\totd M / \totd E_{\rm orb}$ is the distribution of
ejected mass with orbital energy. For orbits with durations
longer than the $r$-process ($\sim 1$~s), the energy deposited
by radioactive heating can exceed the orbital binding energy \citep{Metzger+10a}.
Given the parameters of our simulation, this would occur for
simulation times
\begin{equation}
\label{eq:heating_time_modify}
t \simeq 4 \left(\frac{M}{11.1M_\sun} \right)\left(\frac{2\textrm{ MeV}}{E_r}\right)^{3/2}\textrm{ s},
\end{equation}
where $E_r$ is the energy deposited by the $r$-process. For a range $E_r = 1-3$~MeV,
the affected times are $2-11$~s.

Figure~\ref{f:fallback_dim}a shows the mass accretion rate at the inner
radial boundary of the 3D computational domain ($r = r_{\rm cut}=800$~km) 
for model F3d, which does not include the effect of the disk wind or radioactive heating,
instead simply letting all material evolve under the effects
of gravity. The accreted material at $r = 800$~km is initially composed primarily
of fluid tagged as \emph{fallback}. Around $t=2$~s, the composition
becomes dominated by \emph{tidal tail} material. As pointed out in \S\ref{s:initial_condition},
approximately $0.01M_\sun$ of tidal tail material is gravitationally bound
due to spatial overlap with other components. This material is
almost completely accreted during the simulated time.

The accretion rate at $r=r_{\rm cut}$ in all models that 
do not include the disk wind is shown Figure~\ref{f:fallback_dim}b.
Models F2d and F3d (no radioactive heating) differ slightly 
in their initial evolution until $\sim 1$~s, after which they
display a nearly identical accretion history. The fact that the
time dependence of this accretion is very nearly $t^{-5/3}$
indicates that the mass distribution is close to uniform
in orbital energy, and that the contribution from the fluid pressure
to the dynamics is minor.

\begin{figure}
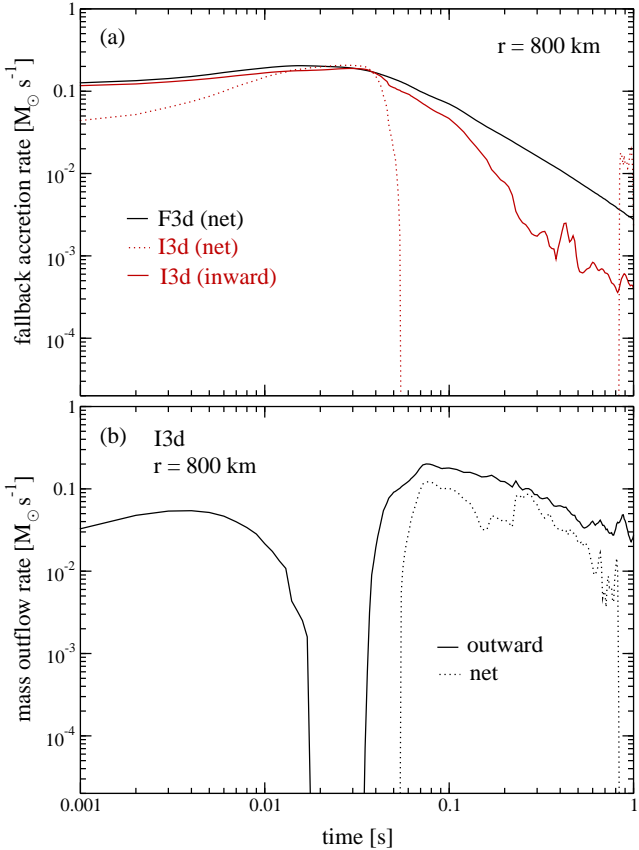

\begin{overpic}[width=\columnwidth,clip=true]{f6a.eps}
\put(3.5,47){\tiny $\odot$}
\end{overpic}
\begin{overpic}[width=\columnwidth,clip=true]{f6b.eps}
\put(3.5,52){\tiny $\odot$}
\end{overpic}
\caption{\emph{Panel (a)}: accretion rate at $r=r_{\rm cut}=800$~km in models
that evolve the dynamical ejecta in 3D with wind injection (I3d) and without 
wind (F3d). The curve labeled `inward'
includes only material initially present in the domain and with
negative radial velocity at the inner boundary, for comparison with the model without wind. 
The net accretion rate for model I3d includes all material.
\emph{Panel (b)}: Mass outflow rate in the fiducial model I3d. Shown 
are the total amount of material with positive velocity injected into
the domain (`outward') and the net outflow rate including all material.
The first outflow episode ($t < 0.02$~s) corresponds to dynamical ejecta material 
moving outward, with the second caused by the disk wind (c.f. Figure~\ref{f:mdot_largeradius}).}
\label{f:fallback_time}
\end{figure}

Including radioactive heating modifies the evolution 
of the accretion rate on the expected timescales 
(eq.~[\ref{eq:heating_time_modify}]) by steepening
the time-dependence over the interval $0.3-3$~s. 
This is shown in Figure~\ref{f:fallback_dim}b, where models 
F2d-h and F3d-h are shown alongside models that do not 
include heating. While a gap in the fallback rate,
as envisioned by \citet{Metzger+10a}, does not appear, the
accretion rate is suppressed over a finite interval
relative to the case without heating, returning
later to the approximate $t^{-5/3}$ scaling. The absence of a gap can
be explained by the constancy of $dM/dE_{\rm orb}$
(equation~\ref{eq:mdot_fallback_definition}) as inferred
from the models without radioactive heating. Addition
of energy by the $r$-process simply shifts mass in this distribution
towards higher energies, filling the gap near $E_{\rm orb} = 0$
with material that initially had lower energy.

\begin{figure}
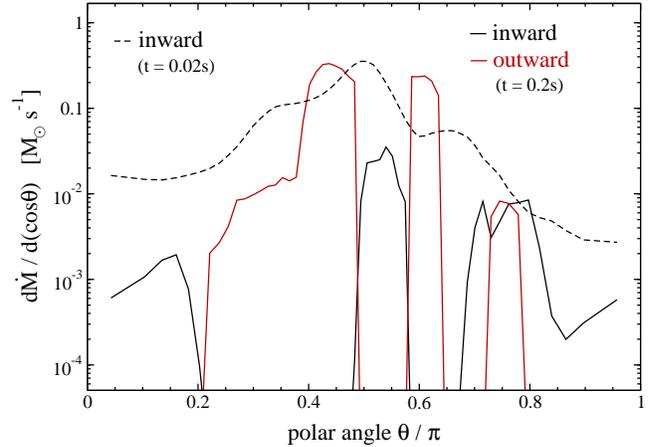

\begin{overpic}[width=\columnwidth,clip=true]{f7.eps}
\put(3.5,49){{\tiny $\odot$}}
\end{overpic}
\caption{Distribution of the mass flow rate at $r=r_{\rm cut}$ 
as a function of polar angle (eq.~\ref{eq:mdot_angle}) for the fiducial 3D model 
with wind injection (I3d) at two times. 
The inward accretion rate includes only material initially present in the domain
and which has negative radial velocity at the inner radial boundary. The outflow rate includes
only material injected into the domain. Compare with Fig.~\ref{f:fallback_time}.}
\label{f:mdot_angle}
\end{figure}

The total accreted material for all models in this sequence
lies in the range $0.026-0.029M_\sun$.
If the accretion rate were to continue indefinitely with the
same magnitude and scaling as it has at $t = 10$~s, only
an additional $\sim 10^{-4}M_\sun$ would be added. 

\subsection{Effect of disk wind on fallback accretion}

\begin{figure*}
\begin{overpic}[width=\columnwidth,clip=true]{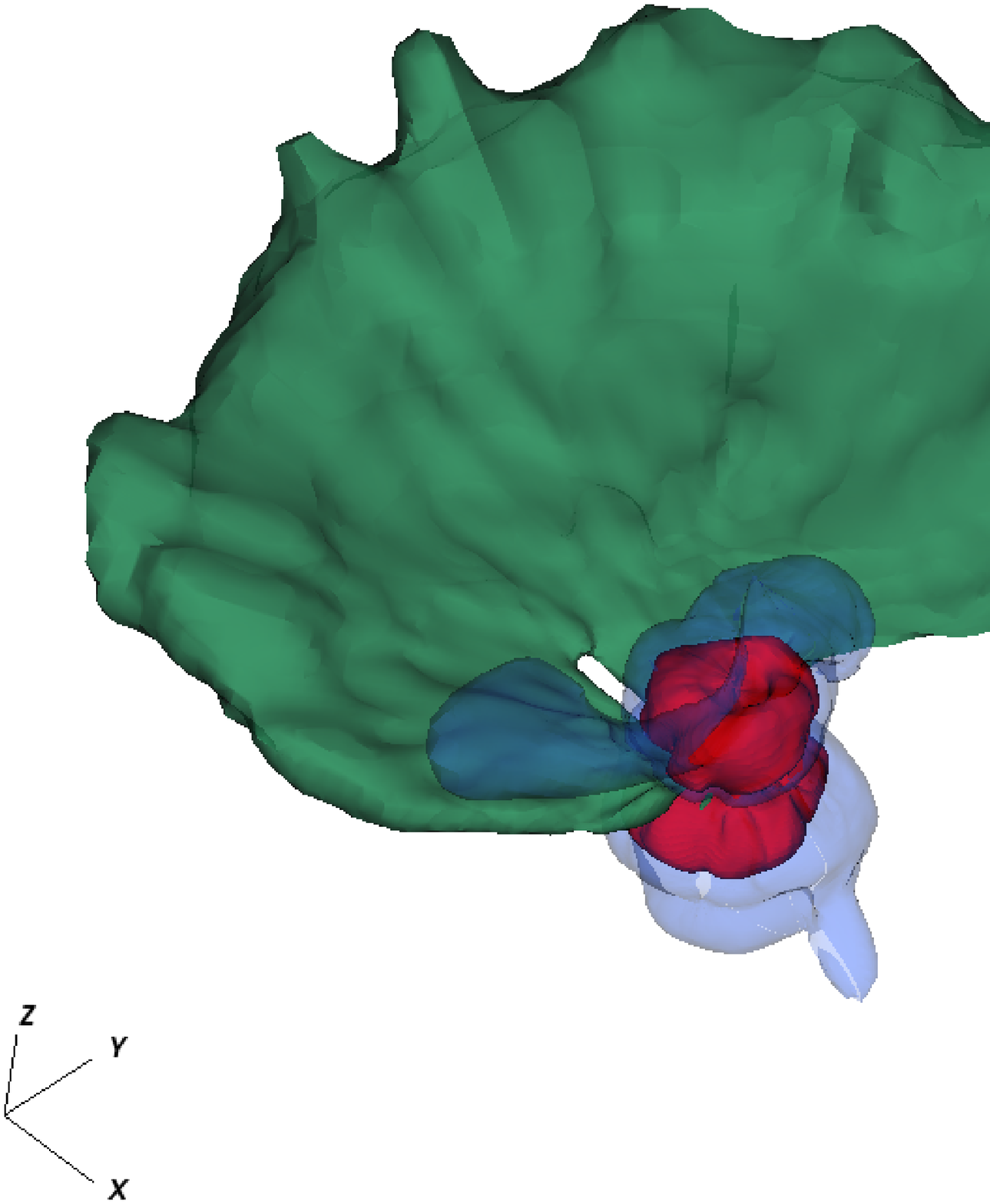}
\put(10,80){{\color[rgb]{0,0.75,0} tidal tail}}
\put(17,77){{\color[rgb]{0,0.75,0} \vector(1,-1){7}}}
\put(78,20){{\color{red} disk wind}}
\put(75,22){{\color{red} \vector(-2,1){7}}}
\put(40,85){no radioactive heating}
\end{overpic}
\begin{overpic}[width=\columnwidth,clip=true]{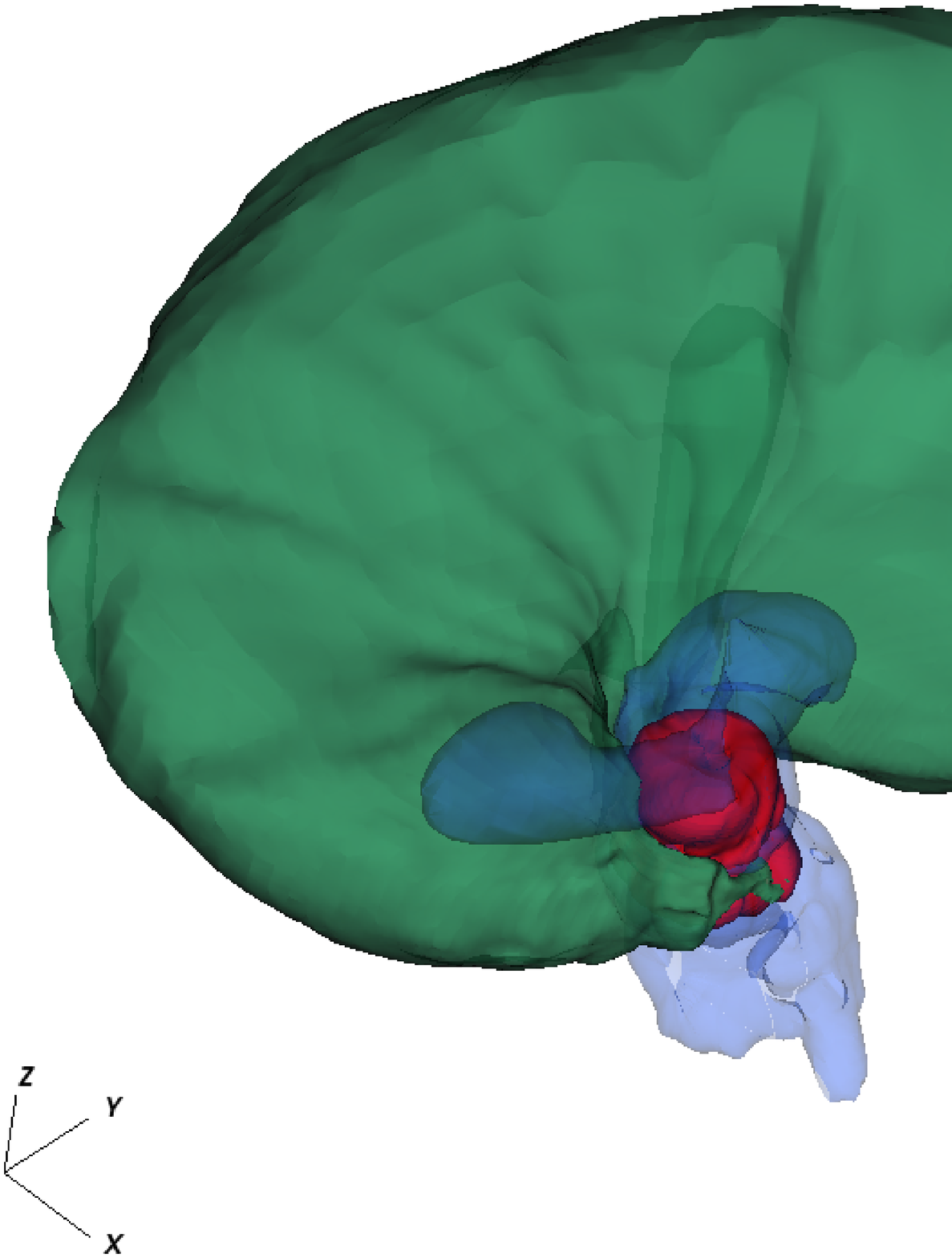}
\put(10,80){{\color[rgb]{0,0.75,0} tidal tail}}
\put(15,77){{\color[rgb]{0,0.75,0} \vector(1,-1){7}}}
\put(78,20){{\color{red} disk wind}}
\put(75,22){{\color{red} \vector(-2,1){7}}}
\put(85,10){\line(1,0){9.2}}
\put(85,10){\line(-1,0){9.2}}
\put(76,5){{\small $5\times 10^{10}$~cm}}
\end{overpic}
\caption{Isosurfaces of passive scalars tracing material initially tagged as 
\emph{tidal tail} (90\% mass fraction, green) and \emph{wind (disk+fallback)} 
(5\% and 95\% mass fraction, blue and red, respectively)
at time $t=10$~s. Shown are the fiducial 3D model of dynamical ejecta evolution with disk
wind injection from the inner boundary (I3d, left) and a version that adds radioactive
heating by the $r$-process (I3d-h, right). Most
of the material shown is already in homologous expansion, hence its geometry 
will not change at later times.}
\label{f:tail_wind_3d}
\end{figure*}

The disk wind causes important changes in the
properties of fallback accretion. This was already seen in the 2D
results of \S\ref{s:2d_evolution}, where the disk completely dominates over
fallback material in setting the late-time accretion rate (Fig.~\ref{f:mdot_isco}).
Here we examine this interplay using more realistic 3D simulations, in which
the wind measured in 2D models is injected from the inner radial
boundary at $r=r_{\rm cut}=800$~km.
Figure~\ref{f:fallback_time}a compares the evolution of the 
mass accretion rate in our 3D simulations
with and without wind. 
We isolate material that is initially outside $r=r_{\rm cut}$
by assigning a passive scalar $X_{\rm inj}=1$
to all material that is subsequently injected into the domain, independent
of which component of the system (\S\ref{s:initial_condition})
it is made of. We then compute the total mass flux with
negative radial velocity at the inner boundary, multiply by $(1 - X_{\rm inj})$, and
integrate in solid angle.
Figure~\ref{f:fallback_time} shows that fallback accretion
is suppressed after $t \sim 100$~ms when the wind is injected. 

Small quantitative modifications in the evolution of the
accretion rate around and after the onset of the wind
are obtained when including radioactive heating, when using a different treatment for 
the wind injection, and when using the wind from model C2d-df, which does
not include the feedback from the tidal tail in 2D (c.f. Table~\ref{t:models}). 

The net accretion rate (all material) at $r=r_{\rm cut}$ for the fiducial model
is also shown in Figure~\ref{f:fallback_time}a. Initially, this
net accretion rate is lower than the case with no wind injection.
This decrease is caused by dynamical ejecta material moving outward,
as shown in Figure~\ref{f:fallback_time}b. Once this initial
outflow subsides, the net accretion rate
reaches its full magnitude around $t\sim 30$~ms. Shortly
thereafter, the net accretion rate becomes net outflow
once the wind turns on.

Figure~\ref{f:mdot_angle} illustrates the simultaneous 
flow of wind material and fallback accretion at $r=r_{\rm cut}$.
Shown are snapshots of the mass inflow and outflow
rate as a function of polar angle,
\begin{equation}
\label{eq:mdot_angle}
\frac{\totd\dot{M}}{\totd(\cos\theta)} = \int\totd\phi\,r_{\rm cut}^2\, f_{\rm inj} \rho v_r,
\end{equation}
with $f_{\rm inj} = X_{\rm inj}$ for outward moving
material ($v_r >0$) and $f_{\rm inj} = (1-X_{\rm inj})$ for $v_r<0$.
At $t = 0.2$~s, accretion proceeds primarily along the equator, with the
wind flowing towards mid-latitudes. This segregation is
not persistent, however, with a different distribution in angle
at later times. Note also that accretion and wind can co-exist
at different azimuthal angles (e.g. $\theta \simeq 3\pi/4$).

At about $t\sim 1$~s, our diagnostic for the inflow 
at $r=r_{\rm cut}$ 
shows that net accretion resumes,
with strong stochastic fluctuations (Fig.~\ref{f:fallback_time}a).
This is, however, related to the spreading of the accretion disk
outside $r=800$~km. In other words, the outer edge
of the radiatively-inefficient, convective accretion 
disk enters the 3D computational domain, with the
wind-launching radius moving continuously outward\footnote{While the position
of the outer edge of the disk is not uniquely defined, we estimate it
by computing an isodensity surface at $10^{-4}$ times the instantaneous density
maximum.}.

\begin{figure*}
\includegraphics*[width=\textwidth]{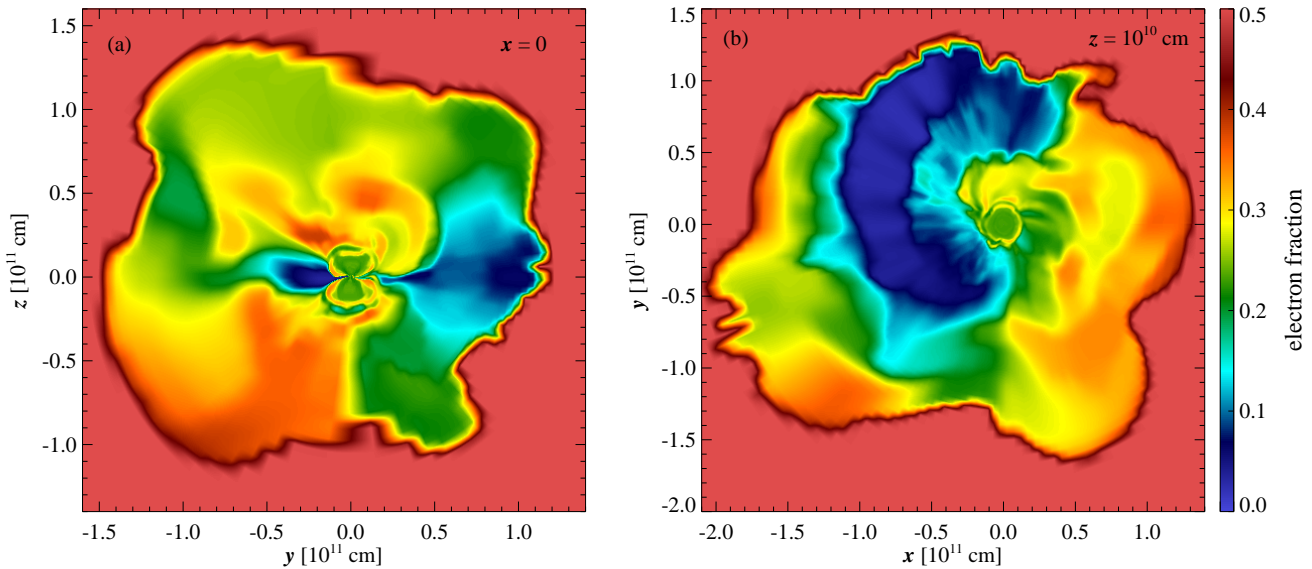}
\caption{Electron fraction in the fiducial 3D model I3d at time $t=10$~s. 
Shown are two-dimensional slices
normal to the $x$-axis (a) and normal to the $z$-axis (b). Compare with Figure~\ref{f:tail_wind_3d}.}
\label{f:ye_slices}
\end{figure*}

\subsection{Properties of the disk wind and dynamical ejecta at large radii}

By the end of our simulations, at $t=10$~s, the system is approaching
homology: the material outside $r\simeq 2\times 10^{10}$~cm has radial velocity
roughly proportional to radius ($\sim r^{0.95}$). This material amounts to
$60$ percent of the mass in the computational domain.

The geometry of this homologous ejecta is shown in Figure~\ref{f:tail_wind_3d}. 
The tidal tail wraps around the rotation axis, occupying primarily the equatorial plane.
The wind is located at the center of the domain, expanding towards high- to intermediate
latitudes. Because the wind is expanding more slowly than the dynamical ejecta by a factor of
several, in the homologous limit its spatial extent must be smaller by the same factor.

Including radioactive heating smoothes inhomogeneities in the tidal tail,
increasing its vertical extent due to the added thermal energy, as shown
in Figure~\ref{f:tail_wind_3d}. This result
is consistent with the findings of \citet{rosswog2014}. In terms of the 
expected electromagnetic counterpart, this implies that neutron-rich 
(and hence high-opacity) material in the tidal tail
obscures the disk wind component for a larger set of viewing angles 
along the equator (e.g., \citealt{kasen2014}), relative to the case without 
radioactive heating. Given this particular set of initial conditions, however, the 
tidal tail does not cover all azimuthal angles, hence the expected optical
emission from the disk wind can readily escape along those unobstructed
viewing directions. 

\begin{figure}
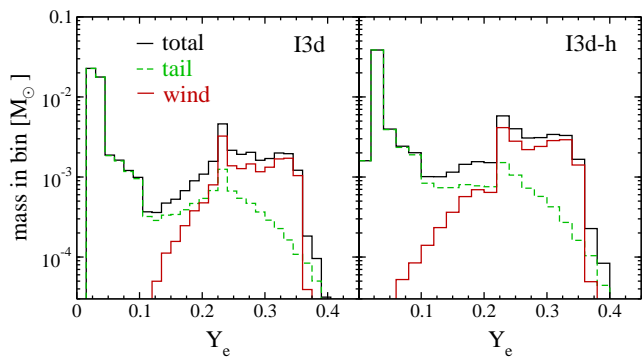

\begin{overpic}[width=\columnwidth,clip=true]{f10.eps}
\put(2,41.3){{\tiny $\odot$}}
\end{overpic}
\caption{Mass histograms as a function of electron fraction for models
I3d and I3d-h (c.f. Figure~\ref{f:ye_slices}). The histograms are computed
by summing all the material crossing the radius $r=3\times 10^9$~cm over
the entire simulation time ($t=10$~s).}
\label{f:histogram_Ye}
\end{figure}

The material in the tidal tail is significantly more neutron 
rich than that in the disk wind, as is well-known. Figure~\ref{f:ye_slices} 
shows slices of the electron fraction distribution normal to the 
$x$ and $z$ axes, illustrating the spatial distribution of material
that will give rise to Lanthanide-rich ($Y_e \lesssim 0.25$) and
Lanthanide-poor ejecta ($Y_e \gtrsim 0.25$; see, e.g., \citealt{kasen2014}).
Figure~\ref{f:histogram_Ye} shows a mass histogram as a function
of electron fraction for models I3d and I3d-h, including all material crossing a surface 
at $r=3\times 10^9$~cm. The histograms are bimodal, with clear contributions
from the tidal tail at $Y_e \lesssim 0.05$, and disk wind at $Y_e \sim 0.25$.
At small amount of tidal tail material is mixed with the wind, and
has higher electron fraction. 

Our method of injecting the wind from the inner radial boundary works
well as long as the disk does not enter the computational domain.
Once the disk enters, around $t \sim 1$~s, there is a discrepancy
between the stresses at this boundary and those that would be 
obtained in a self-consistent simulation. In particular, the use
of an outflow boundary condition whenever the radial velocity
at the boundary is negative leads on average to lower pressure
support on the section of the disk that has entered the domain.

The consequence of this discrepancy in stresses is a decrease in
the amount of mass ejected to large radius in models with wind
injection relative to a self-consistent simulation. Table~\ref{t:models}
shows that this discrepancy is a factor of $\sim 2$, and is
independent of whether 2D or 3D is employed (model I2d
vs. C2d) or whether neutrino and viscous source terms are included
in the self-consistent model (models C2d vs. C2d-src).

We can nevertheless still compare the bulk properties of the wind
at large radius between 3D models, using the 2D model with wind injection (I2d)
as a baseline. Table~\ref{t:models}
shows that the specific wind injection method (solving a Riemann problem or
simply filling the ghost cells with the sampled wind) is largely
unimportant in determining the wind properties at a radius $r=10^9$~cm. Including radioactive
heating does indeed lead to more mass ejection, with an enhancement similar
to that observed in the 2D models. Similarly, injecting the wind sampled
from model C2d-d (no tidal tail) leads to correspondingly small mass ejection.

The velocity and electron fraction of the wind undergo small changes relative to 2D. 
While both increase relative to the 2D models, the change is not likely to lead to qualitative 
differences in their nucleosynthetic properties and in their effect
on the observed kilonova. In the case of the electron fraction, this
implies that there is no significant mixing between wind material and
the bulk of the dynamical ejecta, which would otherwise have lowered $Y_e$ from its
pure wind value. While Figure~\ref{f:tail_wind_3d} indicates that some
mixing does indeed occur in the immediate vicinity of the wind, the amount
of mass affected is a small fraction of the total, and goes in the direction
of making tidal tail material more proton rich (Figure~\ref{f:histogram_Ye}). 
This low degree of mixing is a consequence of the factor $2-4$ faster radial 
velocity of the dynamical ejecta.

\section{Summary and Discussion}
\label{s:summary}

We have investigated the interaction between the disk
wind and dynamical ejecta generated in a NS - BH merger.
Starting from the output of a Newtonian merger simulation, we have 
disentangled the ejecta components using its
phase space distribution (\S\ref{s:initial_condition}). The disk is located at
small radii and is nearly axisymmetric, hence the resulting
disk wind can be estimated from axisymmetric simulations to first
approximation. Given that viscous and neutrino source terms
are sub-dominant in the outer regions of the system, where
the non-axisymmetric dynamical ejecta resides, one can evolve this component
in 3D without the high computational cost of viscous or neutrino
processes. We therefore inject the axisymmetric disk wind 
from the inner boundary of the 3D computational domain. 

By following this two-step approach, we obtain the following results:
\newline

\noindent
1. -- Fallback accretion can be suppressed once the disk wind is
launched (Fig.~\ref{f:fallback_time}). In our models, this
happens $\sim 100$~ms into the simulation. 
\newline

\noindent
2. -- The properties of the disk wind are not significantly affected 
by the dynamical ejecta. This is largely due to the difference
in expansion velocities.
Most of the gravitationally
bound part of the dynamical ejecta (`fallback', \S\ref{s:initial_condition})
is swept up by the wind, forming its leading edge (Fig.~\ref{f:wind_2d_phases}d).
While some small amount of mixing occurs between material coming from the disk and
that in the tidal tail, it goes primarily in the direction of making tidal
tail material more proton rich (Fig.~\ref{f:histogram_Ye}).
\newline

\noindent
3. -- The time dependence of the mass accretion rate at the ISCO steepens after
the wind is launched. For our choice of parameters, it follows a slope
$\sim t^{-2.2}$ (Fig.~\ref{f:mdot_isco}). This is nearly 
independent of whether the \emph{tidal tail} and \emph{fallback} 
components are included in the simulation. This power-law decline is set by the
physics of the viscously spreading disk with outflows \citep{metzger2008steady},
not fallback.
\newline

\noindent
4. -- We do not find a gap in the fallback accretion rate induced by radioactive
heating when the wind is ignored (Fig.~\ref{f:fallback_dim}). Instead,
we find a steepening in the time-dependence of this accretion rate over
a finite interval, resuming the quasi-ballistic $t^{-5/3}$ time-dependence
at late times. Inclusion of radioactive heating in disk wind simulations
yields a small ($\sim 10\%$) enhancement in the ejected mass (Table~\ref{t:models}), 
with no qualitative differences in the ejecta composition (Table~\ref{t:wind_2d}).
\newline

\noindent
5. -- We find that including radioactive heating smoothes out inhomogeneities
in the tidal tail (Fig.~\ref{f:tail_wind_3d}), in agreement with \citet{rosswog2014}.
This results in very neutron-rich material (with high optical opacity) obscuring a larger 
fraction of the available viewing directions towards the wind ejecta,
which has a smaller size in the homologous limit due to its smaller velocity.
However, the specific model we evolved is such that the tidal tail does not
cover all azimuthal angles (Fig.~\ref{f:ye_slices}), leaving a wide range of
unobstructed sight lines. 
\newline

\noindent
6. -- The properties of the disk outflow
are qualitatively similar to those obtained from simulations
with spinning BHs that
begin from idealized initial conditions with somewhat less
massive and more compact disks (e.g., \citealt{FKMQ14}).
\newline

While our quantitative results are specific to the particular initial
condition we have adopted, our approach contains several features
that can be useful in future studies of the interplay between the
different components of the combined merger ejecta. First,
we have found a novel way to isolate the dynamical ejecta -- bound
and unbound gravitationally -- from rotationally supported material.
Second, we have shown that ignoring neutrino and viscous source
terms at large radii ($r\gtrsim 10^8$cm) has a minor effect
on the evolution of the system. The main drawback of our
method is the treatment of the boundary 
condition when the radial velocity is negative. At times $t\gtrsim 1$s, 
the disk has spread out sufficiently to enter our 3D domain.
If the stresses ensuing from this boundary condition are not
sufficient to support the part of the disk inside the domain, 
an excess of mass will flow inwards,
significantly affecting the amount of wind launched to large radii.
Large-domain axisymmetric simulations are a reasonable way to improve 
this analysis, given the similarity in the globally integrated
properties.

\begin{figure}
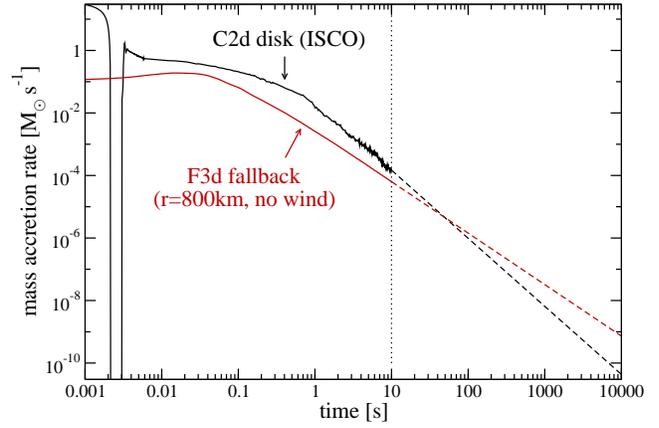

\begin{overpic}[width=\columnwidth,clip=true]{f11.eps}
\put(3.5,49.5){{\tiny $\odot$}}
\end{overpic}
\caption{Extrapolation of the mass accretion rate from disk
accretion at the ISCO (black, model C2d) and fallback with no wind 
at $r=800$~km (red, model F3d) after $t=10$~s, illustrating the smaller energy 
released by the disk at times where extended GRB central enging
emission and/or X-ray flares takes place.}
\label{f:mdot_fallback_extrapolation}
\end{figure}

The fact that we have found no significant mixing between the disk
wind and dynamical ejecta means that separately estimating
the nucleosynthetic contribution from the disk and tidal tail is a reasonable
approximation. This is relevant given that the two components
are expected to lead to different nucleosynthetic signatures, 
with implications for the dispersion in the $r$-process abundance
in the galaxy \citep{just2014} and the properties of the
kilonova emission (\citealt{MF14}, \citealt{kasen2014}).

Even though our results show that fallback accretion is not always
a robust source of late-time engine activity in short GRBs, the fact
that accretion from the disk continues for a long time
provides a persistent source of accretion power. However, the steeper
decline with time in the accretion rate relative to 
ballistic fallback ($t^{-2.2}$ vs. $t^{-5/3}$) 
implies that after $10^4$~s, the energy output from accretion is 
$\sim 30$ times smaller (disk accretion is $\sim 3$ times larger
than wind-free fallback at $t=1$ second, 
cf. Figs.~\ref{f:mdot_isco} and \ref{f:fallback_dim}). This is
illustrated in Figure~\ref{f:mdot_fallback_extrapolation}, which
shows the extrapolation of the mass accretion rate at the ISCO for
the default 2D model C2d, together with the extrapolation of the
fallback accretion rate (without the effect of the disk wind) 
at $r=800$~km from model F3d. The extrapolation of the fallback
rate is an upper limit, however, given that material has angular
momentum and may circularize at a radius larger than the
ISCO. In this case, material will contribute to whatever remains of
the disk and accretion will proceed at a rate set by viscous processes, 
eventually acquiring a $t^{-2.2}$ time dependence.
Even if fallback material has low angular momentum, the presence of the disk
will prevent it from falling directly onto the BH, particularly 
along the midplane.

If accretion indeed dominates late-time engine activity, 
time-variability can result if instabilities 
occur in the outer disk \citep{perna2006} or
near the black hole due to magnetic effects \citep{proga2006}.
Even if accretion is smooth, late-time variability 
can result if the surrounding medium is excavated by Poynting flux
if the neutron star is a pulsar before
the merger \citep{holcomb2014}. 

The observed suppression of fallback in our models is contingent on the
initial condition we have chosen to carry out our study, and hence it is not
necessarily a general property of NS-NS or NS-BH mergers. In our simulations, 
the amount of mass ejection in the 
wind ($0.04M_\sun$, or $\sim 20\%$ of the initial disk mass) is
larger than the initial amount of fallback and bound tidal tail
material ($0.03M_\sun$). Inclusion of general relativity and
a slightly different set of binary parameters (including eccentricity; \citealt{East+12}) 
can in principle lead to a different hierarchy. For example, \citet{foucart2014} find
disk masses in the range $0.04-0.14M_\sun$ and bound dynamical 
ejecta in the range $0.03-0.05$ when considering
mergers of neutron stars with $7M_\sun$ BHs in general relativity.
Their lowest disk mass is a factor $5$ smaller than ours, and the 
expected fallback is a factor $\sim 2$ larger. 
In this case, does the disk wind escape at intermediate latitudes 
while fallback proceeds and keeps the
mass supply constant? Or does fallback suppress the onset of
the wind, entraining material back to the disk?

In the case of NS-NS mergers, the dynamical ejecta is 
less concentrated in the midplane than in BH-NS mergers of
large mass ratio (e.g. \citealt{bauswein2013,Hotokezaka+13}). 
For similar relative masses between bound dynamical ejecta and disk,
the more spherical geometry should make it easier for the disk wind
to disrupt fallback.

Our calculations can be improved in many ways. 
Injection of the wind into an expanding boundary in 3D models
would alleviate the problems introduced when the disk
enters the domain, and allow a better estimate of the
degree of mixing between tidal tail and disk.
The wind calculations can be made more realistic by including MHD and
GR effects self-consistently. The composition of the wind can be better 
quantified by using an improved neutrino transport scheme. 
The contribution of radioactive heating to the dynamics of fallback
material can be studied further by accounting for the increase in temperature
of fluid elements as they fall towards the BH (the prescription we employed
assumes that all fluid elements are in continuous expansion, thus overestimating
the energy release).
Future studies will address these improvements.

\section*{Acknowledgments}

We thank Brian Metzger, Frank Timmes, and the anonymous referee for constructive
comments that improved the paper.
RF acknowledges support from the University of California Office of the President, and
from NSF grant AST-1206097.
EQ was supported by NSF grant AST-1206097, the David and Lucile Packard Foundation,
and a Simons Investigator Award from the Simons Foundation.
JS is supported by the National Science Foundation Graduate Research Fellowship 
Program under Grant No. DGE 1106400.
DK was supported in part by a Department of Energy Office of Nuclear Physics Early 
Career Award, and by the Director, Office of Energy Research, Office of High Energy and 
Nuclear Physics, Divisions of Nuclear Physics, of the U.S. Department of Energy under 
Contract No. DE-AC02-05CH11231.
SR was supported by the Deutsche Forschungsgemeinschaft (DFG) under 
grant RO-3399, AOBJ-584282 and by the Swedish Research Council (VR) under grant 621-2012-4870.
The software used in this work was in part developed by the DOE NNSA-ASC OASCR Flash Center at the
University of Chicago.
This research used resources of the National Energy Research Scientific Computing
Center (NERSC), which is supported by the Office of Science of the U.S. Department of Energy
under Contract No. DE-AC02-05CH11231. Computations were performed at
\emph{Carver} and \emph{Hopper} (repos m1186, m1896, and m2058).

\appendix

\bibliographystyle{mn2e}
\bibliography{ms}

\label{lastpage}
\end{document}